# Workplace sustainability or financial resilience? Composite-financial resilience index

**Elham Daadmehr[1]**

**Abstract**

Due to the variety of corporate risks in turmoil markets and the consequent financial distress especially in COVID-19 time, this paper investigates corporate resilience and compares different types of resilience that can be potential sources of heterogeneity in firms' implied rate of return. Specifically, the novelty is not only to quantify firms' financial resilience but also to compare it with workplace resilience which matters more in the COVID-19 era. The study prepares several pieces of evidence of the necessity and insufficiency of these two main types of resilience by comparing earnings expectations and implied discount rates of high- and low-resilience firms. Particularly, results present evidence of the possible amplification of workplace resilience by the financial status of firms in the COVID-19 era. The paper proposes a novel composite-financial resilience index as a potential measure for disaster risk that significantly and persistently reveals low-resilience characteristics of firms and resilience-heterogeneity in implied discount rates.

**JEL Classification** C11 · C38 · G11 · G12 · Q51 · Q54

Warm thanks to Marco Pagano, Lorenzo Pandolfi and Giovanni Walter Puopolo. I acknowledge participants at the International Risk Management Conference-IRMC2023, the Global Finance Conference-GFA2023, LSC conference "Sustainability and firm performance in Europe and the Americas", Leibniz Institute for East and Southeast European, NSEF workshop at the University of Naples Federico II (UNINA) and the Centre for Studies in Economics and Finance (CSEF) for their comments. Special gratitude to the UNINA for providing research facilities on data and to the Vienna University of Economics and Business (WU-VGSF) for visiting opportunities in 2021 and providing I/B/E/S forecasts data from Refinitiv-Eikon (Thomson Reuters).

✉ Elham Daadmehr
  elham.daadmehr@unina.it

1    Department of Economics and Statistics, University of Naples Federico II, Via Cintia, Monte S. Angelo, 80126 Napoli, Italy

## Introduction

The worldwide economic crisis caused by COVID-19 is systematically different from past crises. The COVID-19 crisis was caused by a viral pandemic that severely shrank global economic activities and the consequences are completely different from previous crises with respect to the cause, scope, and severity. Since 2020, the emergence of COVID-19 and investors' public awareness about its consequences prepares a natural experiment to study the effect of this kind of shock on price fluctuations and firms' discount rates. The COVID-19 pandemic has triggered enormous and heterogeneous stock price movements.

Different firms may exhibit different reactions to the shock and see different expected returns. The goal of this paper is to investigate the importance of corporate resilience. Since resilience is closely related to the functionality and valuation of firms in the COVID situation, this study proposes a measure of corporate resilience and shows whether this characteristic can be a new source of risk and variation in expected return proxies.

In general, resilience is the ability of firms to survive in crisis and to effectively recover during and after an unexpected shock. On the one side, due to the intrinsic feature of the health crisis caused by COVID-19 that immediately hit the workforce, firms with fewer employees experienced a severe decline in employment (Papanikolaou and Schmidt 2022), or firms with higher workforce and communication flexibility would have less managerial issues under new social distancing rules and lockdowns. On the other side, firms could increase the chance of survival from banks to increase their liquidity. Pagano and Zechner (2022) interpret the possibility of the increase in firms' liquidity buffers not only through bond issuance and cash-preserving policies but also through the availability of government grants. As a consequence, firms can become more resilient during the COVID-19 crisis. These two aspects show the necessity of a hybrid definition of resilience.

Since there is a lack of common vocabulary and definition of resilience, this research clarifies the concepts of resilience considered in this paper and compares two types of resilience, to propose an appropriate meaningful hybrid measure for not only in the fever period of the COVID-19 pandemic but for all periods before and after the onset to capture different aspects of resilience.

First, the workplace resilience index is introduced based on what Koren and Peto (2020) propose for firms' flexibility towards social distancing rules. Social distancing restrictions and other mitigation rules against the COVID-19 pandemic hit businesses, especially those that depend on human interactions. These reductions in face-to-face activities directly affect firms' production costs (Koren and Peto 2020). Such disruptions from new social distancing rules can be directly related to new firms' characteristics showing the degree of workplace flexibility and resilience in these situations.

Second, the heterogeneous effect of COVID-19 shock can be amplified by the financial status of companies (Ramelli and Wagner 2020). The financial strength of firms is a key element that can potentially magnify the impact of workplace resilience. Firms with an appropriate level of financial ratios can potentially overcome the side effects of social distancing rules during COVID-19 despite some resilience inconveniences in their workforce. Pagano and Zechner (2022) provide various evidence regarding the financial impact of COVID-19 shock on firms. They mention the moderating role of capital markets and corporate financial policies, specifically, they characterize a firm's performance based on net sales and total assets which could better capture the effect of COVID-19 on their balance sheets. Meanwhile, the provided credits, especially for small firms, can affect the capital structure and increase the leverage, since many of these firms were compensated by the Fed's packages, e.g. Paycheck Protection Program (PPP) for small businesses. The provided evidence by Fahlenbrach et al. (2021) asserts the impact of corporate leverage and cash holdings on the cross-sectional variation of U.S. firms' stock prices in the COVID-19 situation. Thus, it is totally hard to understand how firms financially respond to COVID-19 shock. This paper defines the 'financial-based resilience' index as a representative measure of the financial status of firms, constructed in a screening procedure based on all firms' corporate financials, and emphasizes an important indirect role of main financial ratios on overall firms' resilience. The noteworthy feature is directly related to the 'machine-based mechanism' that chooses the most effective financial ratios for both before the onset of COVID-19 and the period after that. Moreover, the proposed methodology here introduces profitability, valuation, and liquidity as the most effective financial ratios. So, this index provides an opportunity not only to capture the effect of total assets showing the immediate impact of COVID-19 on a firm's investment decisions but also to consider the possible effect of fiscal support policies by different measures of profitability ratios. The results accord with Pagano and Zechner (2022) who mention the usage of large liquidity cushions from banks and financial intermediaries to improve firms' balance sheets in order to be more resilient to the following COVID-19 consequences. The adopted methodology exhibits that among all kinds of firms' financial ratios, operating profit margin, price-operating earnings and quick ratio play key roles in the characterization of firms' financial-based resilience. Specifically, in line with Acharya and Steffen (2020) who explain the role of liquidity on firms' performance in the first quarter of 2020, the financial-based resilience index emphasizes that the importance of liquidity (absolute amount of estimated loading) increases from before to after the onset of COVID-19, as opposed to the price-operating earnings (as a valuation ratio). The proposed machine-based characterization shows a good extent of consistency with Glossner et al. (2020) findings, especially where they emphasize the relation between valuation and the amplification of severe price decline in COVID-19.

As a final proposed measure, the 'composite-financial resilience index' is introduced to exhibit firms' resilience ability not only regarding their workplace flexibility but also representing their financial strength. This paper defines the composite-financial resilience index based on these two intuitions that measures firm's resilience and emphasizes the importance of financial resilience. It proposes composite-financial resilience index as an appropriate measure to provide evidence showing to what extent the impact of workplace resilience can be potentially amplified by financial intuition of resilience. The explanation is provided by comparing the implied discount rates of high- and low-resilience groups of firms.

To do this, the paper considers the simple present value model where the implied discount rate equates discounted expected future cash flows to current prices (Landier and Thesmar 2020). Due to the importance of analysts' forecasts and the reported recommendations (Kothari et al. 2016), this paper considers analysts' earnings expectations as a firm-level proxy of expected future cash flows. It provides evidence of the strong explanatory power of analysts' expectations (in line with De la and Myers 2020; Bordalo et al. 2020) and its important role in the resilience-heterogeneous implied discount rate, especially through composite-financial definition of resilience. The evidence on earnings expectations reveals that analysts are still pessimistic about the low workplace-resilience firms with weaker corporate financials and emphasizes to what extent not only earnings expectations but also implied discount rates are resilience-heterogeneous across firms. Results reveal the inadequacy of considering just one of these two intuitions of resilience as a resilience measure and meanwhile, establish the necessity of both workplace resilience and financial-based resilience in significantly detecting the resilience-heterogeneity.

The novelty of this study is to present an appropriate index for firms' financial resilience and moreover is a unique study that proposes directly an index to capture such aggregate impact of these two kinds of resilience attributed to the COVID-19 era. Finally, it shows that the composite-financial resilience index improves the two former measures of resilience and can 'significantly' and 'persistently' reveal low-resilience characteristics of firms, those that are riskier during all periods of the COVID-19 crisis.

In order to reach the aim, after the literature and data description in "Further related literature" and "Data" sections, respectively, this paper emphasizes the necessity of financial intuition of resilience and clarifies the definition of resilience indexes in "Resilience indexes" section. It exhibits the appropriateness of the proposed hybrid resilience index using earnings expectations in "Discount rate" section where it also reveals the significant resilience-heterogeneity in firms' implied discount rates and suggests the appropriate index of resilience; Then concludes.

## Further related literature

The paper contributes to plenty of literature in resilience studies which has recently attracted loads of attention in economics and finance. Resilience is the capacity of a firm to absorb disturbance (Walker et al. 2004) and to retain the same level of functionality. One approach to resilience is related to the status of some predefined economic variables as structural measures of firms that interpret their performance. For instance, when the economy sees an oil price shock, consumption and income are the most popular benchmarks of the firms' resilience (Sedlmayer and Boehm 2011; Accorigi 2011; Bunting et al. 2011). In finance, Hua et al. (2020) introduce a measure of resilience based on liquidity and its relation to expected returns.

Resilience implicitly includes the cost of change in performance during the crisis to incorporate the idea that the nature of the crisis affects the interpretation of resilience and measurements. In other words, the type of crisis and the corresponding shock determine the definition and the role of resilience; Firms may be resilient to one type of shock but not to other types. The main contribution of this research is to measure corporate resilience that can reveal the effect of COVID-19 on the heterogeneity in stock expected returns.

In the COVID era, the economic concept of performance should be related to environmental aspects caused by COVID, specifically, it should represent to what extent firms' performance depends on communication restrictions and social distancing rules. Since January 2020, there have been many studies about the effect of COVID-19 on the labor market due to the effect of the lockdown and new restrictions on the labor force. Many of them try to propose a measure of workplace flexibility. Koren and Peto (2020) examine business disruption and provide a theory-based measure for the dependency of US businesses on human interaction. They explain how face-to-face communication affects the cost of producing goods and introduces the average 'affected share' based on different social distancing indexes. They interpret that a higher affected share of firms implies less flexibility towards social distancing restrictions. In a separate study, Dingel and Neiman (2020) examine the economic effect of social distancing measures. They conduct surveys on a sample of US workers to find out how many jobs can be done from home. The proposed occupation-level social-distancing measure is constructed using pre-pandemic data. Meanwhile, they examine the positive correlation of these data with the data obtained in the early stage of the pandemic and explain that workplace-flexible jobs pay more than inflexible jobs. They interpret how this measure can vary across industries and countries. Another study by Hensvik et al. (2020) investigates to what extent the COVID-19 pandemic affects the number of job searches and amplifies the labor demand shock. They analyze the intensity and direction of searches in the early COVID crisis and explain the impact of job seekers. Using the diff-in-diff method, they sort the most resilient and the least resilient occupations and individual vacancies that belong to occupations for which labor demand is more resilient to the crisis. This paper takes into account workplace resilience based on Koren and Peto (2020) in order to capture the impact of COVID-like shock and social distancing rules and improves the methodology by other financial aspects of resilience.

Moreover, during the market distress in the early 2020, the study of the relationship between corporate characteristics and market drop conducted by Ding et al. (2021) starts to answer the question that which firm characteristics make some companies more 'immune' than others to the COVID-19 shock. They show that firms with better pre-COVID finances saw a less severe drop in their stock prices. Furthermore, Ramelli and Wagner (2020) identify corporate debt and cash holdings as important drivers of firms' value. They mention the key role of leverage and cash holdings in firm valuation. Specifically, less-levered nonfinancial firms with higher cash are less affected by COVID-19 (Glossner et al. 2020). All of these[1] motivate to define resilience as one of the main characteristics of firms not only in the sense of workplace flexibility that is necessary due to the special type of disaster in the pandemic era but also from the financial resilience perspective. This paper mostly contributes to the recent literature on corporate resilience and its novelty is to present a financial hybrid resilience index to capture such effects of these two kinds of resilience related to the COVID-19 era. In the methodology of the proposed composite-financial resilience index, this paper contributes to the application of cutting-edge literature on the multivariate functional principal component analysis presented by Happ and Greven (2018) and Happ-Kurz (2020).

---

[1] Together with what is explained in "Introduction" section.

On the other side, among huge stock market literature, Alfaro et al. (2020) explain the role of unanticipated infection changes on forecasts of aggregate equity market return and Albuquerque et al. (2020) discuss how abnormal returns react to ESG ratings. As mentioned above, many studies shed light that firms' financial status can be potentially important to overcome the economic consequences of social distancing restrictions and lockdowns. Pagano et al. (2023) demonstrate that among all firms those that are less affected by social distancing restrictions have higher returns in time of the COVID crisis. Meanwhile, Cheema-Fox et al. (2020) show that more flexible firms in terms of workforce, especially those that protect supply chains during the market decline, have higher returns. This paper contributes to a fast-growing strand of literature on COVID-disaster risk and heterogeneous expected return in a different way. It provides a comparison between the three introduced resilience indexes to see to what extent each of these indexes can reveal significant resilience heterogeneity in expected returns. In pricing methodology, the literature mostly introduces different types of present value models based on a wide range of assumptions. Among all studies, Botosan and Plumlee (2005) assess five proxies for firm-specific cost of equity capital. They find that the association between the target price method and PEG ratio method, and the firm-risk measures accords with theory. They explain how this method can be derived from the present value model and the model of economy-wide growth presented by Ohlson and Juettner-Nauroth (2005) by assuming no growth in abnormal earnings beyond the forecast horizon. In the COVID-19 situation, Landier and Thesmar (2020) suggest that discount rate shocks are the main driver of the V-shaped evolution of stock prices. They use analysts' expectations as forecasts of future cash flows and show the term structure of equity discount rates is not flat which is an important component of firm valuation.

In line with Landier and Thesmar (2020), Engelberg et al. (2018), Lochster and Tetlock (2020); and De la and Myers (2020), who discuss the effect of cash flow expectation movements on medium-term price fluctuation, this research considers earnings expectations to obtain implied discount rates, and follows present value model of Landier and Thesmar (2020) and the Gordon growth model for beyond the forecast horizon. Then, it is possible to show to what extent each index can significantly exhibit resilience-heterogeneity in implied discount rates and to see whether the riskier firms can be recognized based on the proposed composite-financial resilience index.

## Data

In order to figure out if the proposed hybrid index can significantly reveal a sign of resilience risk, resilience indexes as well as implied DR should be calculated. This requires a wide range of different data. The following is a short explanation to clarify all these types of data.

This paper is going to introduce three resilience indexes in the following section: workplace resilience, financial-based resilience, and composite-financial resilience. For the workplace resilience index, this paper uses the proxy presented by Koren and Peto (2020), namely, 'affected share'. This proxy is proposed based on the early period of COVID-19 observations in the first quarter of 2020 and computed for each 3-digit NAICS industry. About the full explanation of 'affected share', this paper directly refers to the original paper; Meanwhile, "Resilience indexes" section provides a brief explanation of this theory-based measure.

To construct financial-intuition resilience indexes, the computation procedure uses more than 70 monthly financial ratios of nearly 5833 U.S. firms from 2013 to 2022, including all categories: capitalization, efficiency, profitability, liquidity, solvency, valuation, and financial soundness, available at WRDS database. Table 2 in Appendix provides the exact definition of all these financial ratios.

For earnings expectations, daily consensus analysts' earnings (I/B/E/S forecasts) is used to figure out the annualized growth of these expectations for all U.S. firms with high or low workplace flexibility, from 15 February to 11 May for the current (2020) and three following fiscal years. These earnings expectations are available at Refinitiv-Eikon, Thomson Reuters. To calculate implied discount rates, monthly consensus analysts' earnings of all U.S. firms is considered in 2020 as a current fiscal year and four following fiscal years from I/B/E/S-WRDS.

To complete the computation of implied discount rates, the paper uses closing price over 2020–2022 and fundamentals: dividends, common stock repurchases from 2010 to 2019, and net income from 2015 to 2019, available at Compustat/CRSP merged database for the U.S firms. "Discount rate" section explains the pricing methodology in detail.

## Resilience indexes

### KP workplace resilience index

Due to the nature of the COVID-19 pandemic shock and the importance of defining the theory-based measure of resilience, this study firstly considers the intuition of resilience based on Koren and Peto (2020). Their methodology reveals the impact of COVID-19 social distancing restrictions on the economic functionality of firms.

In principle, Koren and Peto (2020) describe social distancing measure as a link between the effect of COVID restrictions on human interaction and the economic consequences on firms' performance. They define three dimensions of occupations: teamwork-intensive, physical presence, and customer-facing and introduce a theory-based measure that explains to what extent businesses rely on face-to-face communication (Charlot and Duranton 2004 and Tian 2019) and which firms are mainly affected by these restrictions. Their model of communication shows the sensitivity of production costs to an increase in face-to-face interaction and determines firms that perform less efficiently from home.

In their model, the firm's production process is in terms of the number of workers and the firm optimally decides on task sharing between workers. In a cost minimization problem, they define a cost function of firms consisting of total communication and production costs and find the optimum number of workers. Furthermore, they define the social distancing effect caused by new COVID restrictions which reduces face-to-face contact to an exogenous threshold amount of N. They conclude that production costs increase with social distancing when the optimal number of workers is greater than N. They introduce 'affected share' referring to the share of workers in any of the three occupation groups. This study defines the KP workplace resilience index as equivalent to 'affected share'; And the higher amount of affected share implies the lower resilience (Koren and Peto 2020).

By definition, the KP index shows how this measure explains the degree of firm's compatibility with lockdown and newly legislated rules and emphasizes the economic performance of firms. Based on Koren and Peto (2020), low-resilience firms are not economically sustainable since they encounter business disruption more than other firms, especially due to restrictions on human interaction. Thus, it is expected to see such firms to be more sensitive to new social distancing restrictions and lockdowns. In other words, the KP index of resilience measures the economic sustainability related to 'workplace' flexibility, apart from to what extent the firm was financially strong before the COVID-19 pandemic.

The important feature of the KP index is that the resilience of firms depends on their own workplace characteristics and flexibility towards the new mitigation policies which is not implied by the workplace-resilience of other firms. Despite all the prominent features of this resilience measure, "Discount rate" section shows the shortcoming of this index to exhibit 'significantly' the resilience-heterogeneity in the firm's implied discount rates and proposes how it is possible to overcome the issue.

Figure 1 reveals the heterogeneity of workplace resilience across industries and provides a relative comparison on the frequency of different sectors with the same degree of resilience for the U.S. firms.[2] This figure asserts that most firms categorized as high workplace resilience firms, specifically in 'Manufacturing: Iron and steel equipment, agricultural machinery, and electrical and computer products' sector, whereas it is expected not to be such high resilient, since around 90% of the world economy had been under some form of lockdown, disrupting supply chains, depressing consumer demand and putting millions of job losses. Moreover, the developed economies were expected to contract by 5% in 2020. In the following sections, other resilience indexes are introduced to understand how it is possible to overcome this puzzle.

## Financial-based resilience index

Apart from workplace flexibility as a main characteristic that empowers firms to survive in the COVID-19 era, this subsection presents a hierarchical machine-based mechanism to define a new 'financial-based resilience' index based on all financial ratios, belonging to all categories: Capitalization, efficiency, profitability, liquidity,

---

[2] The sample is based on all U.S. firms with available data for computation of implied discount rate: analysts' earnings expectations, payout ratio, growth rate, etc.

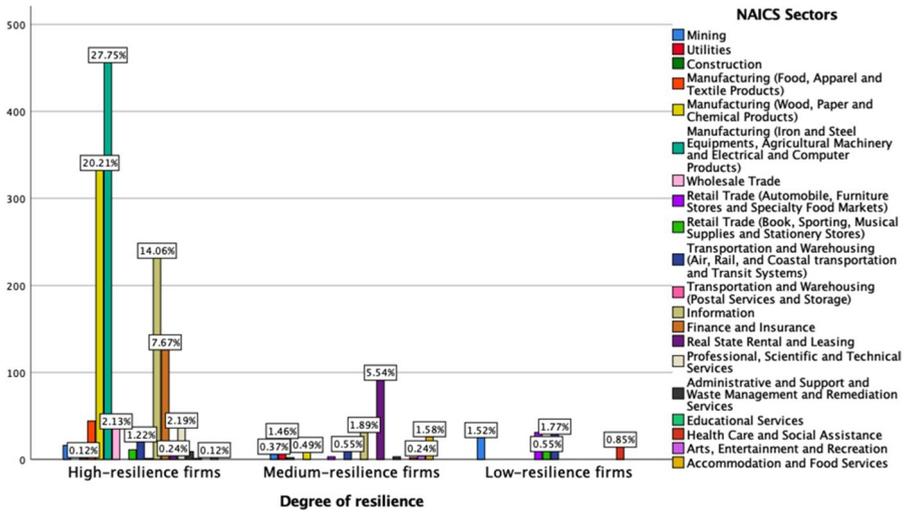

**Fig. 1** The frequency of NAICS sectors in each category of workplace-resilience: firms with affected share less than 40 are assigned to a high-resilience group and ones with greater than 65 are assigned to the low-resilience group. The remainders are considered for the medium-resilience category. *Data source* Koren and Peto (2020)

solvency, valuation, and financial soundness,[3] in order to take into account the impact of corporate financials. Although the literature clearly highlights the necessity of each category, the novel presented methodology in this subsection examines which financial ratios outweigh others.

## The necessity of financial intuition of resilience

Due to the different characteristics of firms and several policies adopted to tackle the financial and economic consequences of COVID-19, it is difficult to figure out how firms' financial status plays an important role in such a chaotic situation.

Li et al. (2021) interpret the booster role of firms' financial characteristics, for instance, sales, return on assets, and profit margins, on firms' corporate culture that potentially affects the firms' real performance. The procedure of constructing the financial-based resilience index includes the impact of different kinds of efficiency ratios (e.g. sales-to-equity and sales-to-working capital). Moreover, it is also possible to see some firms are more sensitive to cash-preserving policies during COVID-19, especially firms with lower profitability and revenue growth (Pettenuzzo et al. 2022). The impact of all profitability ratios (e.g. net profit margin and operating profit margin after depreciation) is covered by the machine-based definition of financial-based resilience.

---

[3] Pagano and Zechner (2022) mention an important relation between company size and financial variables like net sale and total assets. In this paper, the financial-based resilience index provides the overall effect of company size through its interaction impact on different financial soundness and efficiency ratios.

Meanwhile, flexibility in business prospects, jointly with financial flexibility could be an efficient path to reach COVID-resilience. Barry et al. (2022) suggest the importance of financial and investment flexibility based on internal funds and external finance. They explain how firms deal with the related managerial issues by capital spending schemes. Moreover, firms' financial policies and equity-raising activities can potentially alter the asset growth rate and consequently, the impact of COVID-19 on firms' financial status. As Pagano and Zechner (2022) suggest, this could happen through raising capital (e.g. bank credit, corporate bond issuance, and equity issuance), corporate policies (e.g. capital structure and liquidity, and also dividend payout and share repurchase policies) and also public policies (e.g. loan guarantees and PPP that can potentially reduce the possibility of business bankruptcies and increase firm's survival for longer time (Wang et al. 2021) by providing more liquidity. However, all these from different perspectives affect the capital structure and emphasize the necessity of an index in order to reveal the overall 'compounding' impact of all these policies. For instance, the increase in liquidity leads to changes in capital structure and a decrease in productivity due to the adverse effect of loan guarantees (Acharya et al. 2021b; Crouzet and Tourre 2021; Kalemli-Ozcan et al. 2022). The machine-based definition of financial-based resilience index considers all capitalization ratios (e.g. common equity-to-invested capital and long-term debt-to-invested capital) to include the effect of capital-raising, solvency ratios (e.g. total debt-to-equity and total debt-to-capital) to capture the impact on capital structure, and liquidity ratios (e.g. cash ratio and quick ratio). Moreover, it is possible that COVID-19 pushes firms to a huge reduction in dividends to compensate for their capitalization and liquidity (Cejnek et al. 2021). Consistently, this paper considers the dividend payout ratio (together with other valuation ratios) as a complementary financial variable.

In what follows, it can be seen clearly that solvency ratios are not major variables to explain the impact of the financial status of firms, as opposed to profitability ratios. This contributes to Blickle and Santos (2022) who compare the indebtedness and profitability of U.S. listed companies, and in line with Pagano and Zechner (2022) who explain that there is no increase in the level of leverage for listed companies due to capital-raising.

## Financial-based index

The procedure is separately conducted cross-sectionally for two periods of time: before the onset of COVID-19 (Jan 2013–Dec 2019) and after the emergence (Jan 2020–Dec 2021). In each period, the average of each financial ratio is computed over time for each firm.

In the first step, principal component analysis determines some financial ratios as major ratios that could significantly explain the overall variation caused by all 70 financial ratios. Those financial ratios that are highly and significantly correlated with selected principal components are considered for the second step of screening. The 23 selected ratios in the first stage are still from all categories (Fig. 11 in

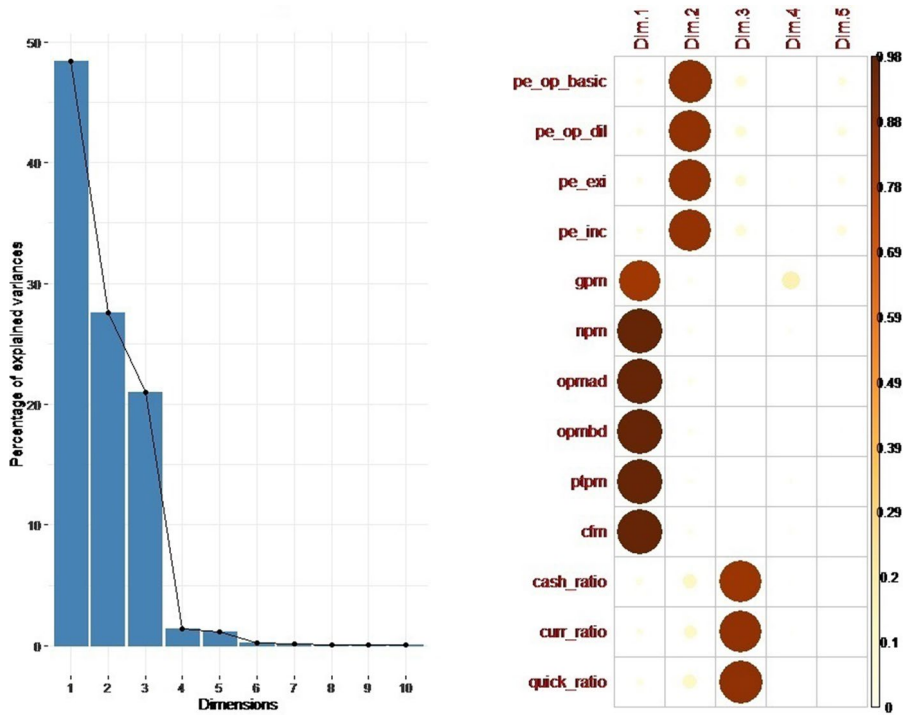

**Fig. 2** Scree and correlation plots: the figure shows results of PCA at the second step before the COVID-19 outbreak. The left panel (scree plot) shows the proportion of variation explained by each component. The right panel (correlation plot) presents the correlation of each financial ratio with the corresponding dimension (component). The size and color of circles represent the correlation between the financial ratios (rows) and principal components (columns). Appendix contains the full explanation of abbreviation names in Table 2. *Data source* Financial ratios, WRDS database. (Color figure online)

Appendix contains the results on the first five components and the corresponding correlations with each of the 23 financial ratios).

In the second step, the procedure determines 13 major ratios (out of the 23 ratios in the first step), belonging to three major components: profitability, valuation, and liquidity. The left panel of Fig. 2 provides statistical evidence that the first three principal components are sufficient to capture the most variation caused by these financial ratios. The right panel shows the correlation of each financial ratio with the corresponding principal components before the onset of COVID-19[4]. It clearly reveals that the first, the second, and the third PCs are related to profitability ratios, valuation ratios, and liquidity ratios, respectively. In each category, the financial ratio with the highest correlation with the corresponding component is considered as the most major financial ratio to define the financial-based resilience index in the third step.

---

[4] Results for after the onset are available in Appendix, Fig. 12.

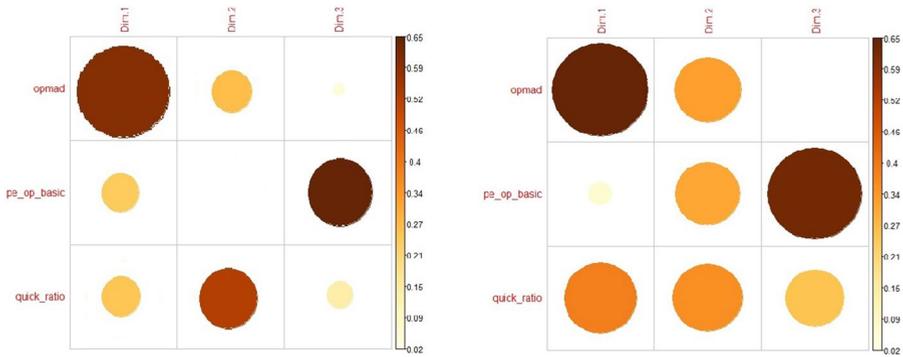

**Fig. 3** Correlation plot: the figure shows the results of PCA at the third step before and after COVID-19. The left panel shows the correlation between ratios and the first three components before COVID-19. The right panel shows the results after COVID-19 emergence. The size and color of circles represent the correlation between the financial ratios (rows) and principal components (columns). Appendix contains the full explanation of abbreviation names in Table 2. *Data source* Financial ratios, WRDS database. (Color figure online)

In the third and final step (Fig. 3), the financial-based resilience index is defined as the first principal component of remainder variables: operating profit margin (profitability ratio), price-operating earnings (valuation ratio) and quick ratio (liquidity). The index is obtained for both periods, before and after the emergence of COVID-19 in the left and right panels, respectively. Table 1 compares the correlation of three selected profitability, valuation, and liquidity ratios with the first component, those that all are significant major contributors to variation originated by all financial ratios and represent the firm's financial status. As it can be seen clearly, profitability and valuation ratios have significant positive loading, as opposed to liquidity ratio.

Financial-based resilience index for firm $i$ at time $t$ is defined as:

$$FB_{it} = \begin{cases} 0.75 opmad_{it} + 0.45 pe_{it} - 0.46 quickratio_{it} & \text{Before COVID-19} \\ 0.75 opmad_{it} + 0.21 pe_{it} - 0.55 quickratio_{it} & \text{After COVID-19} \end{cases}$$

where $opmad_{it}$, $pe_{it}$ and $quickratio_{it}$ stand for operating profit margin, price-operating earnings and quick ratio, respectively.[5] This machine-based specification is in line with Li et al. (2021) who emphasize that profit margins and sales per employee imply strong corporate culture, especially in the COVID-19 era and consequently, mitigate the effect of COVID-19 shock on asset price fluctuation. Moreover, the important role of quick ratio as the major liquidity ratio accords with Pagano and Zechner (2022) who mention the significant change in liquidity of listed U.S. firms. The statistical tests on the level of quick ratio prove the significance of COVID-19's impact on the liquidity level from before to after the onset.[6]

---

[5] The results of statistical tests determine the significant change not only in financial-based index but also in liquidity and valuation ratios from before to after the emergence of COVID-19.

[6] The results are available upon request.

**Table 1** Summary statistics, correlation, and significance of selected profitability, valuation, and liquidity ratios with the first corresponding PC in each period, before and after the onset. *Data source* Financial ratios, WRDS database

| COVID period | Financial ratio | Min | 1st Qu. | Median | Mean | 3rd Qu. | Max | ST.D | Correlation with first PC |
|---|---|---|---|---|---|---|---|---|---|
| Before | Operating profit margin | − 2100.97 | − 0.61 | 0.03 | − 22.10 | 0.14 | 0.96 | 146.15 | 0.80 |
| | Price-operating earnings | − 890.62 | − 6.22 | 13.94 | 3.51 | 22.17 | 296.00 | 67.13 | 0.51 |
| | Quick ratio | 0.073 | 1.09 | 1.72 | 3.31 | 3.40 | 104.17 | 4.78 | − 0.52 |
| After | Operating profit margin | − 1616.66 | − 0.50 | 0.01 | − 14.22 | 0.14 | 0.88 | 98.01 | 0.80 |
| | Price-operating earnings | − 287.91 | − 4.62 | 10.56 | 10.62 | 22.13 | 347.69 | 39.10 | 0.25 |
| | Quick ratio | 0.070 | 0.99 | 1.59 | 3.19 | 3.30 | 79.01 | 4.81 | − 0.62 |

The decrease in coefficient of valuation ratio from 0.45 before COVID-19 to 0.21 after the onset and the higher negative loading of liquidity ratio after the onset show that since the emergence of COVID-19, liquidity moderates more the impact of valuation and profitability ratios, possibly due to the impact of Fed's packages, especially for small businesses. Such packages increase firm's liquidity to make them capable enough to overcome the economic consequences of COVID-19 and to compensate for their financial weakness. This can be potentially a sign of low firms' financial resilience and shows that firms with lower financial resilience benefited from either capital raising (e.g. bank credit) or public policies (e.g. Fed's packages) and have increased their resources to a higher level since they severely needed more liquidity. This suggests the negative loadings for the liquidity ratio, consistent with what is obtained above through hierarchical PCA. Generally speaking, this evidence of a negative sign for the coefficient of liquidity shows a penalty for the degree of financial resilience. The amount of penalty equals the 10% ($-0.62 + 0.52 = -0.10$) decrease in the firm's financial-based resilience (10% increase in its absolute correlation with the first principal component).

The machine-based choice of liquidity ratio for financial-based resilience index by PCA and the significant change in its level from before to after COVID-19, together with no choice of capitalization and solvency ratios, suggest that capital-raising through credit and bond market, as Pagano and Zechner (2022) interpret, did not result a prominent change in leverage (solvency ratios), in line with what is expected about the possibility of external equity during the pandemic. The impact of the external equity could be potentially such high that the effect of COVID-19 on capitalization ratio would not be significant.

The key role of the financial-based resilience index is to quantify the impact of corporate financials. The first plus point of the financial-based resilience index is to capture the effect of all major firm's financial ratios and to reveal the impact of the financial status of firms on resilience and consequent price changes. Section "Discount rate" compares different indexes and provides novel evidence of the amplification of workplace resilience by financial-based resilience (Section "Implied discount rates"). Based on the definition of the financial-based resilience index, this happens due to changes in valuation ratios rather than other ratios. The reason is directly related to a higher decrease in correlation with financial-based resilience for price-operating earnings (Table 1). This evidence is in line with Glossner et al. (2020) who emphasize the key role of institutional investors in valuation and the amplification of price crashes. Moreover, to complete their approach, this paper shows in Section "Implied discount rates" that this financial-based resilience index can exhibit the implicit role of firms' financial status in heterogeneous price fluctuations through earnings expectations. The further interpretation of the performance of this index is provided in the following sections where different resilience indexes are compared and the paper presents an evidence of resilience-heterogeneity in earnings expectations and firms' implied discount rates as well.

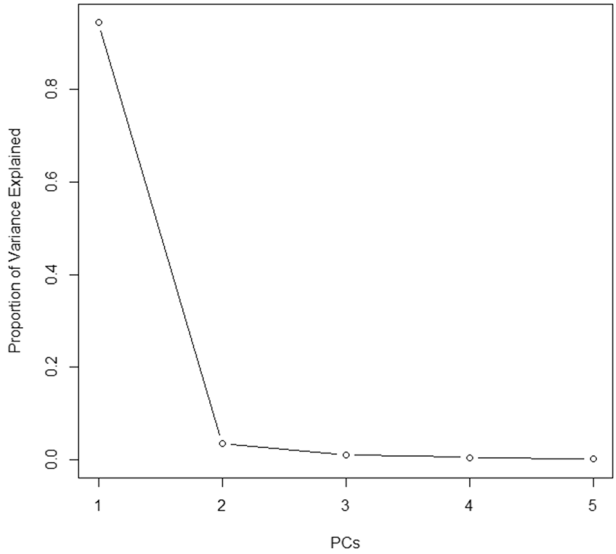

**Fig. 4** Scree plot of multivariate functional principal components: this figure is the result of multivariate principal component analysis of the bivariate of workplace resilience and financial-based resilience. It shows the proportion of variation explained by the first five multivariate functional PCs. *Data source* Koren and Peto (2020) and financial ratios, WRDS

### Composite-financial resilience index

To measure the firm's resilience in COVID-19 crisis, this paper suggests an index based on workplace resilience as a main characteristic linked to the nature of this crisis and financial-based resilience, using a more general framework called multivariate functional principal component analysis.

This method is an application of multivariate Karhunen–Loeve theorem and interprets the $X(t) = (X_1(t), X_2(t))^T = (KP(t), FB(t))^T$, containing workplace resilience (KP) and financial-based resilience (FB), as the following aggregate of infinite eigen-functions, $\psi_m(t)$:

$$X(t) = \sum_{m=1}^{\infty} \rho_m \psi_m(t)$$

where $\rho_m$ is a zero-mean random variable $\rho_m = \langle\langle X(t), \psi_m(t)\rangle\rangle$ with $\mathrm{Cov}(\rho_m, \rho_n) = v_m \delta_{mn}$. Moreover,

$$E\left[\left\|X(t) - \sum_{m=1}^{M} \rho_m \psi_m(t)\right\|^2\right] \to 0 \text{ as } M \to \infty.$$

Then, workplace resilience of each firm, $KP(t)$,[7] is assumed to be affected by zero-mean white-noise over time. The scree plot (Fig. 4) provides the portion of the total variation created by $X(t)$, bivariate of the workplace and financial-based resilience, explained by each multivariate functional principal component and suggests a sufficient number of components ($M$). It shows that the first component explains more than 90% variation of $X(t)$ and $M = 1$ is sufficient to see convergence in the Karhunen–Loeve theorem.

The composite-financial resilience index is defined by $\hat{\rho}_i = \hat{\xi}_i^{(KP)}\hat{c}^{(KP)} + \hat{\xi}_i^{(FB)}\hat{c}^{(FB)}$, with $\hat{c}$ as an eigen-vector of $(N-1)^{-1}\Lambda^T\Lambda$, where $\Lambda_i = (\hat{\xi}_i^{(KP)}, \hat{\xi}_i^{(FB)})$ and $\hat{\xi}_i$ is the corresponding estimated score for firm $i$ based on univariate functional principal component analysis. Intuitively, a decrease in $\hat{\rho}$ is equivalent to less dependency of a bivariate of the firm's workplace and financial resilience to 'time' variation. This implies the lower $\hat{\rho}$ for high-resilience firms since it is expected to see high-resilience firms less affected by crises happening over 'time'.

Following sections show how this index can 'significantly' exhibit the resilience-heterogeneity in firms' cost of capital, and more importantly how it can reveal the amplification impact of corporate financials.

## Discount rate

This section shows to what extent an appropriate resilience index can potentially detect the source of heterogeneity in the implied discount rates as a proxy of expected returns, especially in the COVID-19 era. Practically, it gives the opportunity to investors to assess the risk of an investment and set a benchmark for future investments based on the interest percentage that an investment may yield over its lifetime.

To calculate implied discount rate, this paper follows the present value model and considers discounted earnings expectation (analysts' forecasts) as it can better reveal the firm-level (cash flow) shock rather than future dividend streams. This is a realistic assumption since in the COVID situation, investors become more concerned about earnings and their beliefs are more sensitive to it; Thus analysts' earnings forecasts could be a good representative of expected cash flow by investors. Moreover, this consideration is in line with De la and Myers (2020) who reveal the strong explanatory power of analysts' expectations.[8]

Following Landier and Thesmar (2020):

$$P_{it} = \frac{b_i Ex_t \mathrm{EPS}_{ih}}{1+r_{it}} + \frac{b_i Ex_t \mathrm{EPS}_{i,h+1}}{(1+r_{it})^2} + \frac{b_i Ex_t \mathrm{EPS}_{i,h+2}}{(1+r_{it})^3} + \frac{(1+g_i)b_i Ex_t \mathrm{EPS}_{i,h+2}}{(r_{it}-g_i)(1+r_{it})^3},$$

---

[7] For simplicity, workplace resilience in this section is considered as 100-'affected share', where 'affected share' is introduced by Koren and Peto (2020).

[8] In line with Chen and Zhao (2009) who assert large misspecification in cash flows backed out from discount rates.

where $Ex_t\text{EPS}_{ih}$ stands for consensus earnings expectation of firm $i$ at time $t$ based on the current fiscal year ($h = 2020$ and 2021).[9] $P_{it}$ and $r_{it}$ are closing price and discount rate, respectively. For the payout ratio ($b_i$), first sum of dividends and common stock repurchases is obtained. After normalizing by net income, its average over 2010–2019 is computed and winsorized cross-sectionally at 0 and 1. At each NAICS sector, the average sale growth ($g_i$) over 2015–2019 is scaled by 2015 sales, after omitting outliers.

Following subsections explain the goodness of different resilience indexes based on their ability to categorize firms into high- and low-resilience based on the earnings expectation of firms as well as the corresponding implied discount rates of these firms in the COVID era.

### Earnings expectations

One of the main drivers of heterogeneous discount rates can originate from expected cash flow and earnings expectations due to its ability to capture the effects from the real part of the economy and to reveal the direct effect of COVID-19 on firms' performance. The evidence provided by Li et al. (2021) about the impact of COVID-19 on firm's stock returns based on earnings call transcripts, made reasonable figure out resilience-heterogeneity in earnings expectations. This subsection is going to compare three indexes to see which kinds of corporate resilience are effective on earnings expectations. Using three indexes of resilience explained in the previous section, firms are categorized into three groups and high- and low-resilience firms are compared based on the annualized earnings expectations.

Figure 5 provides a detailed interpretation about the sensitivity of daily annualized growth of earnings expectations in the feverish period of COVID-19 and shows how such forecasted growth evolves negatively for high- and low-resilience firms over time[10], for 2020 and the three following fiscal years. As expected, firms with low-resilience workforce experienced lower growth of earnings expectations, as opposed to the high-resilience ones. Before and during the fever period of COVID-19, low-resilience firms are dominant in the current fiscal year of 2020 (upper-left panel), in the sense that aggregate earnings expectation was affected more by the earnings expectation of low-resilience firms and this dominance effect will diminish in further fiscal years (other three panels).

The annualized earnings expectation of high-resilience firms for the following fiscal years (and similarly the one for low-resilience firms) is shown in Fig. 6. This figure compares the expected growth of these two groups of firms separately for different fiscal years.

Figure 6 reveals that for higher horizons, the growth of earnings expectations is less steep for both high- and low-resilience firms and long-term expectations did not react as much as the short-term one. Figures 5 and 6, suggest that the KP workplace resilience index is a good measure to exhibit the heterogeneity related to workplace

---

[9] The calculation contains earnings expectations for the fiscal year 2023.

[10] Since Koren and Peto (2020) compute 'affected shares' using data of the fever period, categorizing the firms in Figs. 5 and 6 is based on the workplace resilience index, KP.

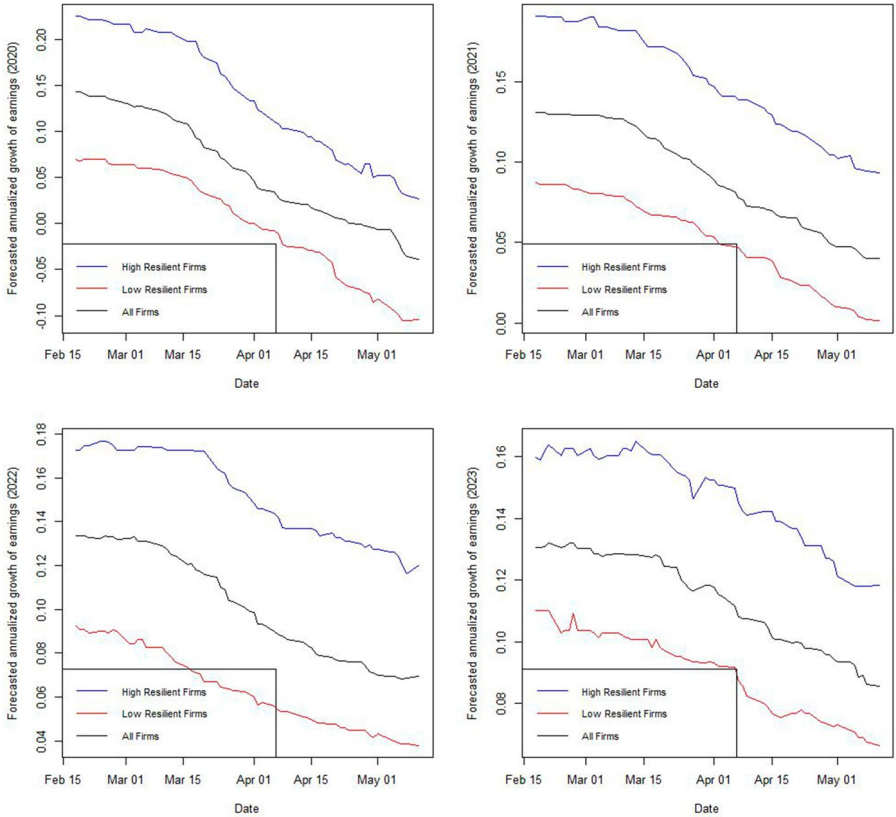

**Fig. 5** Expected annualized growth of earnings: this figure shows the forecasted growth $\frac{1}{h-2019}(Ex_t\text{EPS}_{ih} - \text{EPS}_{i,2019})/\text{EPS}_{i,2019}$ of high- and low-resilience firms based on the KP workplace resilience index during the feverish period for four fiscal years. $Ex_t\text{EPS}_{ih}$ stands for earnings expectation of firm $i$ at time $t$ for horizon $h$. Firms with 'affected share' less than 40 are assigned to the high-resilience group and ones with greater than 65 are assigned to the low-resilience group. *Data source* Koren and Peto ([2020]) and Refinitiv-Eikon platform (Thomson Reuters) daily consensus analysts' earnings (I/B/E/S forecasts)

flexibility during the fever period. This observation is not only in line with Landier and Thesmar ([2020]) but also quite consistent with Koren and Peto ([2020]) and their results on the sensitivity of firms' cost of production to social-distancing restrictions since the onset of the pandemic including the fever period. This emphasizes the necessity of KP workplace intuition as a resilience measure.

Figure 7 shows the overall change in earnings from analysts' point of view for four categories of firms based on their workplace flexibility (KP) and financial-based resilience (FB). The first-left diagram shows that among firms with higher workplace flexibility, those with better financial status see less reduction in earnings expectations from 2020 to 2021. It is also expected to see a steeper rebound in earnings for the following fiscal years (from 2021 to 2024), especially after all Fed's

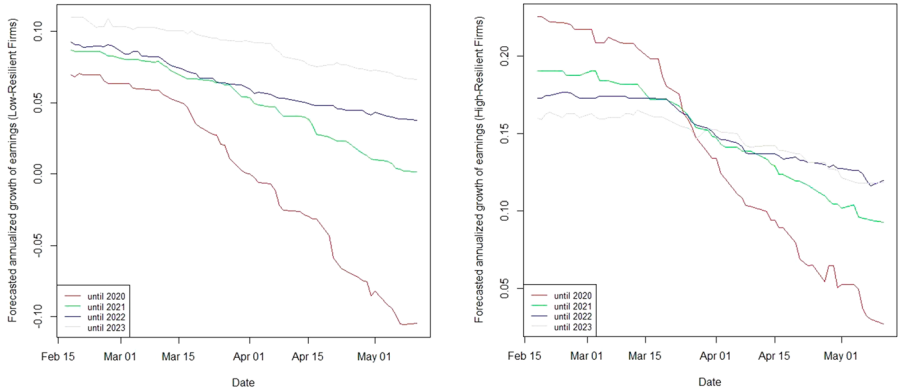

**Fig. 6** Expected annualized growth of earnings of low- and high-resilience firms: the left panel compares the earnings expectations growth $\frac{1}{h-2019}(Ex_t\mathrm{EPS}_{ih} - \mathrm{EPS}_{i,2019})/\mathrm{EPS}_{i,2019}$ of low-resilience firms for higher horizons. $Ex_t\mathrm{EPS}_{ih}$ stands for earnings expectation of firm $i$ at time $t$ for horizon $h$. The right panel prepares similar comparison for high-resilience firms. Firms with 'affected share' less than 40 are assigned to the high-resilience group and ones with greater than 65 are assigned to the low-resilience group. *Data source* Koren and Peto (2020) and Refinitiv-Eikon platform (Thomson Reuters) daily consensus analysts' earnings (I/B/E/S forecasts)

interventions in 2021. This implicitly shows the amplification of workplace resilience by financial-based resilience.

Figure 7 makes the importance of financial-based resilience more clear, specifically its role in expectations of those firms with low flexibility towards COVID-19 social distancing restrictions (Second panel). It clearly suggests that even after all Fed's packages, low-FB firms cannot experience overall upward changes in expectations if they have low degree of workplace flexibility (blue line in the second panel), whereas firms with higher level of financial resilience can continue the upward status quo despite low flexibility in their workplace (the red line in the second panel). Moreover, making a comparison between the first and the second panel reveals the amplification of financial-based resilience by workplace resilience since within either high or low financial-based resilience (high-FB or low-FB) group, higher workplace resilience (high-KP) firms see less reduction in earnings expectations (red (blue) line in the first panel in comparison with a red (blue) line in the second panel). Furthermore, within the low financial-based resilience group (the blue lines in the first and the second panels), high-KP firms are expected to experience a complete rebound after four fiscal years (as opposed to firms with lower workplace resilience). The first and the second panels suggest that the impact of each workplace resilience and financial-based resilience on earnings expectations can be amplified by one another.

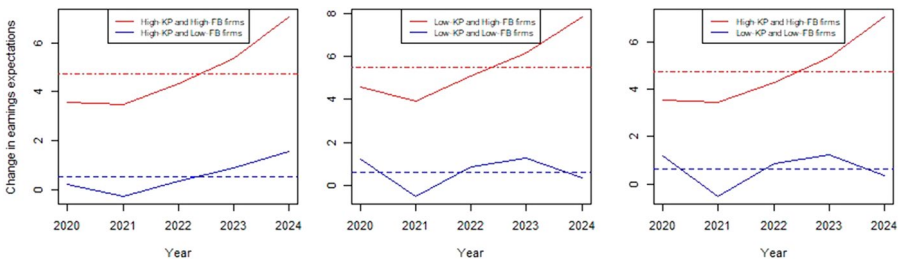

**Fig. 7** Earnings expectations: this figure shows the evolution of an average of $(Ex_t\mathrm{EPS}_{ih} - Ex_t\mathrm{EPS}_{i,\mathrm{Jan}})/Ex_t\mathrm{EPS}_{i,\mathrm{Jan}}$ for four categories of firms based on financial-based resilience and KP resilience over each fiscal year. $Ex_t\mathrm{EPS}_{ih}$ stands for earnings expectation of firm $i$ at time $t$ for horizon $h$. It is obtained for 2020 and four following fiscal years. Based on Koren and Peto (2020), firms with an 'affected share' of less than 40 are assigned to the high-resilience group and ones with greater than 65 are assigned to the low-resilience group. Firms with an average financial-based resilience index higher than the 66th percentile are assumed as high-resilience and firms with lower than the 33rd percentile are the low-resilience ones. *Data source* Koren and Peto (2020) and Monthly consensus analysts' earnings (I/B/E/S-WRDS). (Color figure online)

The third panel shows two extreme cases and exhibits that as opposed to more resilient firms (in the sense of both KP and FB), firms with less workplace flexibility and less financial resilience will generally see downward average revisions in the COVID era.[11] The panel shows that analysts are still pessimistic about the rebound for low-KP and low-FB stocks. This panel emphasizes to what extent earnings expectations can be heterogeneous due to either workplace or financial-based resilience and as a result, suggests to what extent this can affect heterogeneity in firms' implied discount rate. This figure reproves the necessity of the composite-financial resilience index as a hybrid measure of resilience containing both kinds of resilience.

Figure 8 not only reveals the resilience-heterogeneity in analysts' earnings expectations but also makes empirical comparison between the descriptive power of three measures of corporate resilience. This figure shows the overall change in earnings expectations of high- and low-resilience firms categorized based on each resilience index for each fiscal year. It exhibits the inappropriateness of KP and financial-based resilience indexes although the second panel correctly shows that among all firms, those with higher financial-based resilience saw a less severe decline in expectations at the beginning and also a stable trend for the following fiscal years (red line in the second panel). The figure explicitly reveals the important role of workplace resilience on overall composite-financial (CF) resilience since adding the intuition of workplace resilience to financial-based measure gives an improvement in interpretation of earnings expectation growth of low-resilience[12] firms (the blue line in the third panel compared to the one in the first and second panels), in a sense that as it is expected, CF-resilience correctly shows a reduction in these expectations from 2020 to one fiscal year after the onset of pandemic; Meanwhile the effect of adding the intuition of financial-based resilience leads to a correct perception of KP-resilience impact on earnings expectation, specifically for high-resilience firms (the red line in the third panel in comparison with the one in the first and the second panels) in the sense that expectations see less reduction with respect to the one for the low-resilience firms. In general, CF-resilience compares correctly the earnings expectations of these two groups of firms with one another and exhibits more realistically the evolution of overall earnings expectations of high- and low-resilience firms.

---

[11] Revision in the spirit of Landier and Thesmar (2020).

[12] In the sense of multivariate functional principal components and composite-financial resilience.

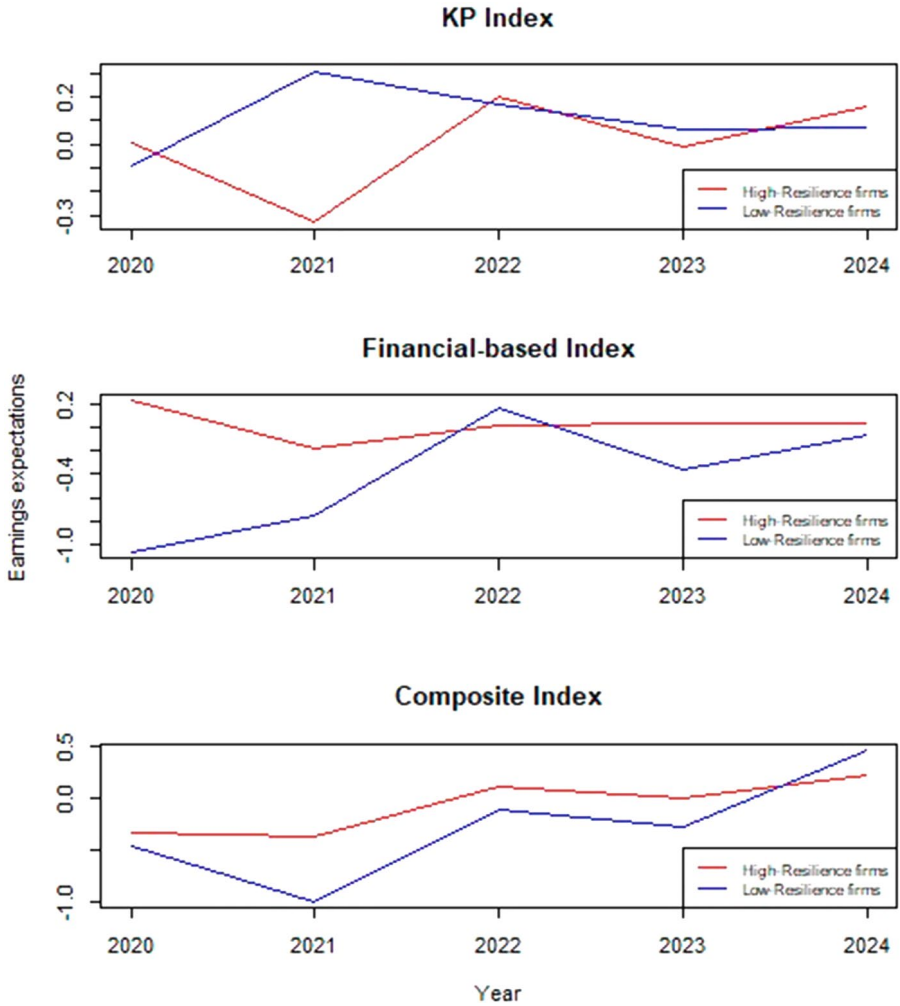

**Fig. 8** Average earnings expectations after COVID emergence: the plot shows the annual average effect of COVID-19 on earnings' expectations of high-resilience and low-resilience firms separately, based on each index of resilience over further fiscal years. This figure illustrates the COVID-19 effect on analysts' expectations originated from the resilience and computed as an annual average of $\frac{Ex_t\text{EPS}_{ih} - Ex_t\text{EPS}_{i,\text{Jan}}}{Ex_t\text{EPS}_{i,\text{Jan}}}$. Based on Koren and Peto (2020), firms with an 'affected share' of less than 40 are assigned to the high-resilience group and ones with greater than 65 are assigned to the low-resilience group. For the second panel, firms with an average financial-based resilience index higher than the 66th percentile are assumed as high-resilience firms and those with lower than the 33rd percentile are the low-resilience ones. Using the composite-financial resilience index, firms with an index less than the 33rd percentile of the composite index are assumed as high-resilience firms and ones with an index higher than the 66th percentile are considered low-resilience ones. The results are robust to categorization using other percentiles. *Data source* Koren and Peto (2020) for workplace resilience, WRDS for firm-level financial ratios and Monthly consensus analysts' earnings (I/B/E/S - WRDS). (Color figure online)

To sum, apart from resilience-heterogeneity in earnings expectations showed by all these indexes, a measure of financial-based resilience is not capable enough to give an appropriate image of the overall corporate resilience regarding to analysts' earnings forecasts. However, this measure together with considering the effect of the social distancing rule on the economic cost of firms improves the understanding of the resilience-heterogeneous earnings expectations. As it can be seen clearly, the third panel emphasizes the importance of this aggregation of resilience intuitions and shows a smoother trend for high-resilience firms in the sense of the composite-financial index and a steeper decline for the low-resilience firms from the emergence of COVID-19 in 2020 to one fiscal year ahead. This exactly refers to the novelty of the new definition, composite-financial resilience index ("Resilience indexes" section).

This strong possibility of the explanatory power of analysts' expectations is in line with De la and Myers (2020). All evidence in this section, in line with Bordalo et al. (2020), suggests that the expectation of earnings is not just backed out of market prices, but is formed independently by analysts, and reflects the heterogeneous expected cash flows. Such resilience-heterogeneity in earnings expectations sheds light on the effect of workplace resilience and financial resilience on firms' expected cash flows and raises a new interesting research idea to model this resilience-heterogeneity which is beyond the scope of this paper. In a separate study, Daadmehr (2022) focuses on this point and proposes a new asset pricing model for COVID-19 time. She shows how the cash flow stream can be affected by the firm's resilience in the time of COVID-19, in line with many studies that emphasize the existence of resilience heterogeneity in dividend stream and the expected returns as well (among all Pagano et al. 2023 and Daadmehr 2023).

In the next part, it is interesting to investigate the footprint of such amplification impact of these two intuitions of resilience and the aggregation of them on heterogeneity in firms' cost of equity capital, as a proxy of expected return, through such heterogeneous earnings expectations; And more importantly, to compare these three resilience indexes.

## Implied discount rates

So far the evidence exhibits how each of resilience indexes is capable of reasonably figuring out the expected future cash flows. This subsection tries to examine if this leads to any evidence of resilience-heterogeneity in the implied discount rate and more importantly, to compare these three resilience indexes. Figure 9 shows the average implied discount rate of high- and low-resilience firms, based on each resilience index. KP workplace resilience index is a strong measure of resilience due to its ability not only to interpret the effect of social distancing on firms' production costs but also to classify firms into two 'more' and 'less' risky categories, as the first panel shows. This observation is consistent with the definition of the KP index as it is proposed based on firms' performance regarding restrictions on human interaction at the onset of the pandemic; However, in what follows the provided evidence shows that the KP measure is less capable in such categorization[13]. The KP index is defined based on firms' production costs that can be related to other financial aspects. What a potentially appropriate resilience index needs is to take into account that firms with better financial status before and during COVID-19 can better overcome to economic consequences of this crisis.

The second panel exhibits that implied discount rate of high financial-resilience firms is smoother and these firms are less risky before the end of the fever period of COVID-19 since they have had better financial status. As time passes they fall into difficulties since some may not be workplace-flexible enough. This panel reveals a flip point in the implied discount rate. Generally speaking, this panel, especially the flip point can be novel evidence of the inappropriateness of this index since it is not possible to reach the unique categorization of high- and low-resilience firms using this measure, meaning that based on this index high-resilience firms are less risky before the end of the fever period but more risky after that. Whereas, high-resilience firms are expected to be less risky not only before but also after the end of the fever period. One may come up with the question of 'How is it possible that low-resilience firms have also a lower implied discount rate than the high-resilience ones, after the feverish period?'. The flip point is in line with the provided evidence by Pagano and Zechner (2022) who mention the importance of small firms (SMEs) in the economy and explain the puzzle that the growth in assets was larger for these firms and so are expected to be less risky but they are not.

As a piece of result, this flip point not only can be originated from the preparation of huge liquidity facilities by the Fed's packages (i.e. moderating role of liquidity) but also can be a sign of consequences of reversal in trading strategies in favor of firms with more flexible workforce after the fever period.

The novel evidence of the flip point suggests a sign of the fear of cash flow dry-up that enforces firms to increase their resources by facilities from banks and capital markets, and taking loans to increase their liquidity. According to the definition of the financial-based resilience index in "Financial-based resilience index" section, this measure captures such injections in the firm's liquidity and the higher injected liquidity is a sign of lower financial-based resilience. Even higher workplace flexible firms in the low-FB category also started increasing their liquidity because of the fears but less than lower workplace resilience firms since they encounter fewer difficulties in their workforce and are more flexible under social distancing rules. Moreover, high-workplace flexible firms are expected to be less risky with lower expected returns as the first panel suggests. Since the financial-based measure of resilience does not take into account the firms' workplace flexibility, firms in the low-resilience category (based on the FB index, in the second panel) can be either high- or low-flexible in their workforce. As a result, it is not surprising to see that low FB-resilient firms have lower implied discount rates and are less risky. If these firms are less risky, it is not because of their financial status but it is due to the workplace flexibility.

This novel evidence shows the necessity of workplace intuition of resilience, in line with Barry et al. (2022) who not only assert that flexibility in financial and investment strategies can potentially help to mitigate the side effects of this crisis but also emphasize the importance of distance-working and workplace flexibility in response to COVID-19 shock. All these show that financial-based resilience cannot be adequate enough to represent the real degree of firms' resilience and as a result, an appropriate resilience index should be a sort of aggregation of these two kinds of index.

---

[13] In the sense that it cannot significantly categorize firms based on their degree of resilience (Fig. 10). The statistical significance is determined by simply comparing the monthly mean of these two groups.

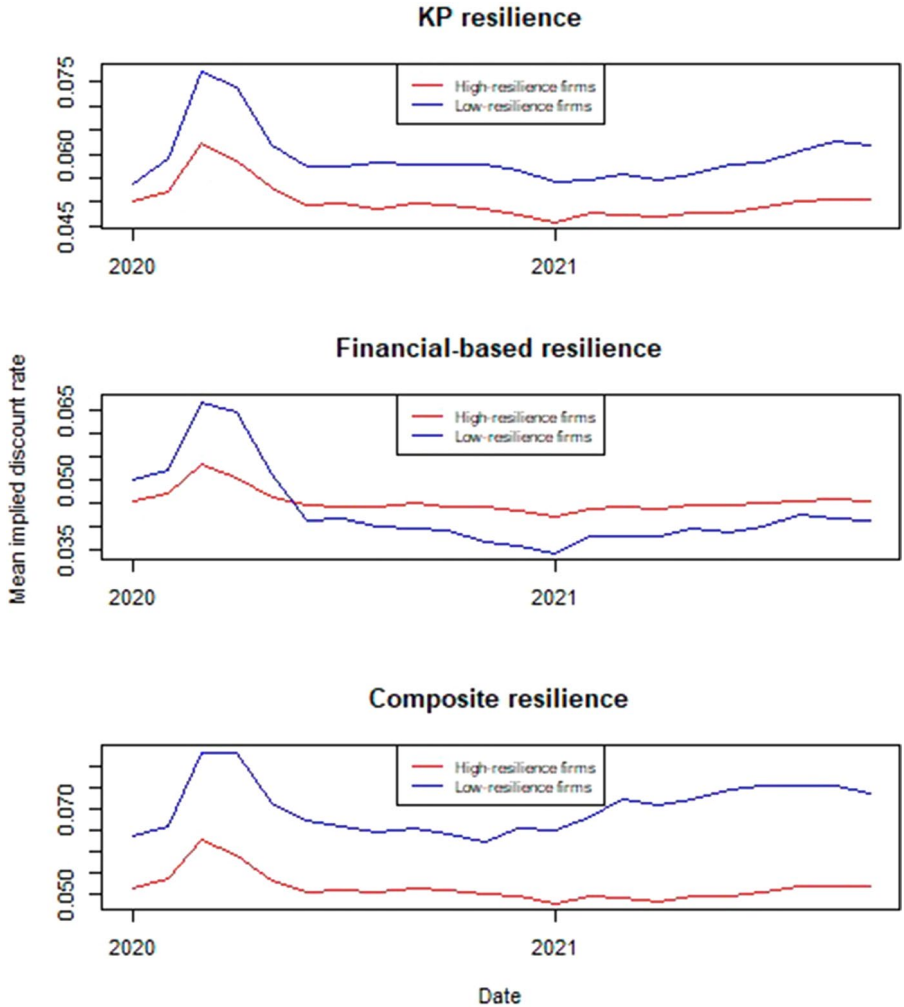

**Fig. 9** The evolution of implied discount rate: The plot shows the average implied discount rate of high-resilience and low-resilience firms separately, based on each index of resilience over 2020 and 2021. The implied discount rate is obtained from the present value model with analysts' earnings expectations as expected future cash flow discounted by payout ratio. For the first panel, firms with an 'affected share' less than 40 are assigned to the high-resilience group and ones with greater than 65 are assigned to the low-resilience group. For the second panel, firms with an average financial-based resilience index higher than the 66th percentile are assumed high-resilience firms and those with lower than the 33rd percentile are low-resilience ones. Using composite-financial resilience index, firms with an index less than the 33rd percentile of the composite index are assumed as high-resilience firms and those with an index higher than the 66th percentile are considered as low-resilience ones. *Data source* Koren and Peto (2020) for workplace resilience, WRDS for firm-level financial ratios, I/B/E/S-WRDS for monthly consensus analysts' earnings and Compustat/CRSP merged for closing price and fundamentals

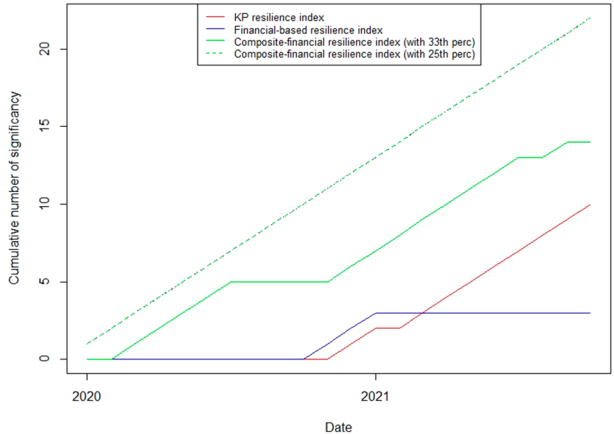

**Fig. 10** The appropriateness of different resilience indexes: this figure plots the results of statistical tests of mean comparison for cross-sectional implied discount rates of high- and low-resilience firms. The black dash-line is a benchmark of goodness (perfect goodness), the case that the average implied discount rates of low-resilience firms are statistically and significantly different from that of high-resilience firms during all this period of time. Other lines correspond to one type of resilience index. The closer to the dash line, the better choice of resilience measure. Based on the KP resilience index, firms with an 'affected share' less than 40 are assigned to the high-resilience group and ones with greater than 65 are assigned to the low-resilience group. Based on the financial-based resilience index, firms with an average of financial-based resilience index higher than the 66th percentile are assumed high-resilience firms and those with lower than the 33rd percentile are low-resilience ones. Using the composite-financial resilience index, the solid (green dash) line is the case that firms with an index less than the 33rd (25th) percentile of the composite index are assumed as high-resilience firms and those with an index higher than the 66th (75th) percentile are considered as low-resilience ones. *Data source* Koren and Peto (2020) for workplace resilience, WRDS for firm-level financial ratios, I/B/E/S-WRDS for monthly consensus analysts' earnings and Compustat/CRSP merged for closing price and fundamentals. (Color figure online)

On the other hand in the third panel, composite-financial resilience index not only presents a smoother implied discount rate for high-resilience firms showing less riskiness (in line with what is implied by financial-based resilience index) but also reveals the amplification impact of workplace flexibility by financial-based resilience over these two years, in the sense that it shows a relatively higher implied discount rate for low-resilience firms than what is obtained by the KP workplace resilience index (the blue line in the third panel compared to the blue line in the first panel), specifically in the fever period, the relative higher peak for low-resilience firms in the feverish period (in the third panel compared to the first panel). All these panels together with the novel evidence of the flip point emphasize the stronger and dominant impact of workplace resilience than the effect of financial-based resilience (in line with Daadmehr 2022).

This directly refers to the novel application of multivariate functional principal component analysis that proposes a sort of method for aggregation of these two kinds of resilience; meanwhile, figures out the novel evidence of amplification. The composite-resilience index in the third panel not only suggests that the implied discount rate of less resilient firms will remain significantly and persistently higher than that of more resilient firms but also shows that the implied discount rate for less resilient firms records the higher level with respect to its initial level, as opposed to the one for more resilient firms (the blue line compared to the red line in the third panel). This suggests that the CF-resilience index is an appropriate resilience measure to detect the longer effect of COVID-19 on the expected return of low-resilience firms.

To sum up, based on the present value model, not only heterogeneity in expected cash flow leads to heterogeneous implied discount rates that one could expect as COVID hit the real part of the economy as well as the stock market but also composite-financial resilience index is the better choice as an index to interpret such resilience-heterogeneity.

Figure 10 shows to what extent the categorization based on each index is statistically significant. This figure shows the monthly mean comparison between the implied discount rates of two groups of firms (high- and low-resilience), categorized based on different resilience indexes, from 2020 to 2021. The black dash-line is a benchmark of goodness (perfect goodness), the case that the average implied discount rates of low-resilience firms are statistically and significantly different from that of high-resilience firms during all this period of time. Other lines correspond to one type of resilience index. The closer to the dash line, the better choice of resilience measure. This figure shows that the composite-financial resilience index can potentially categorize firms significantly into high- and low-resilience risk groups since the implied discount rate for high-resilience firms cross-sectionally significantly differs from the one for low-resilience firms over 2020 and 2021, containing the feverish and the post-COVID era. This plot is empirical proof to show neither the KP nor financial-based resilience index is appropriate enough to significantly reveal the resilience-heterogeneity in implied discount rates. The appropriateness of the composite-financial resilience index is approved for after the Russia-Ukraine war, for the first quarter of 2022, using out-of-sample forecasting.

## Conclusion

This paper aims to propose an appropriate hybrid resilience index as a detector of disaster risk and a new source of heterogeneity in firms' implied discount rates in the pandemic era. It started to set a benchmark to capture the resilience, regarding not only the impact of new social distancing restrictions and lockdown rules but also the financial strength of firms. It provided evidence of the amplification of workplace resilience by firms' financial status.

Using the financial-based (FB) resilience index, it is possible to capture the impact of other effective elements like corporate financials on disaster risk and resilience-heterogeneous implied discount rates. More importantly, the paper provided an interpretation of the flip point in the evolution of the implied discount rate of high- and low-FB resilience firms, after the fever period of COVID-19.

In short, results suggest that financial-based resilience amplifies the impact of social distancing to identify 'significantly' resilience-riskier firms. This explicitly shows the necessity and refers to the novelty of composite-financial (CF) resilience index. The paper employed a multivariate functional principal component analysis to construct and explain this kind of resilience index that can be potentially applicable for the assessment of the pandemic's long-run impacts on firms. The results emphasize the prominent role of the composite-financial resilience index in providing evidence of resilience-heterogeneity in expected payoffs to figure out and better understand how such heterogeneity can transmit to the implied discount rates as well as price changes. This raises a new research idea on asset pricing in the COVID-19 era for future studies (Daadmehr 2022). Results not only suggest that the implied discount rate of less resilient firms, in the sense of 'composite-financial', will remain significantly and persistently higher than that of more resilient firms but also show that the implied discount rate for less resilient firms records the higher level with respect to its initial level, as opposed to the one for more resilient firms. This suggests the longer effect of COVID-19 on the expected return of low-resilience firms that could be investigated more in future research. The datasets generated during and/or analysed during the current study are available from the corresponding author on reasonable request.

## Appendix

See Table 2 and Figs. 11 and 12.

**Table 2** Financial ratios and definitions: *Data source* Financial ratios, WRDS database

| Financial ratio | Variable name | Category | Formula |
|---|---|---|---|
| Capitalization Ratio | capital_ratio | Capitalization | Total Long-term Debt as a fraction of the sum of Total Long-term Debt, Common/Ordinary Equity and Preferred Stock |
| Common Equity/Invested Capital | equityjnvcap | Capitalization | Common Equity as a fraction of Invested Capital |
| Long-term Debt/Invested Capital | debtjnvcap | Capitalization | Long-term Debt as a fraction of Invested Capital |
| Total Debt/Invested Capital | totdebt_invcap | Capitalization | Total Debt (Long-term and Current) as a fraction of Invested Capital |
| Asset Turnover | at_turn | Efficiency | Sales as a fraction of the average Total Assets based on the most recent two periods |
| Inventory Turnover | inv turn | Efficiency | COGS as a fraction of the average Inventories based on the most recent two periods |
| Payables Turnover | pay_turn | Efficiency | COGS and change in Inventories as a fraction of the average of Accounts Payable based on the most recent two periods |
| Receivables Turnover | rect_turn | Efficiency | Sales as a fraction of the average of Accounts Receivables based on the most recent two periods |
| Sales/Stockholders Equity | sale.equity | Efficiency | Sales per dollar of total Stockholders' Equity |
| Sales/Invested Capital | salejnvcap | Efficiency | Sales per dollar of Invested Capital |
| Sales/Working Capital | salejiwc | Efficiency | Sales per dollar of Working Capital, defined as difference between Current Assets and Current Liabilities |
| Inventory/Current Assets | invt_act | Financial Soundness | Inventories as a fraction of Current Assets |
| Receivables/Current Assets | rect_act | Financial Soundness | Accounts Receivables as a fraction of Current Assets |
| Free Cash Flow/Operating Cash Flow | fcf_ocf | Financial Soundness | Free Cash Flow as a fraction of Operating Cash Flow, where Free Cash Flow is defined as the difference between Operating Cash Flow and Capital Expenditures |
| Operating CF/Current Liabilities | ocfjct | Financial Soundness | Operating Cash Flow as a fraction of Current Liabilities |
| Cash Flow/Total Debt | cash_debt | Financial Soundness | Operating Cash Flow as a fraction of Total Debt |
| Cash Balance/Total Liabilities | cashjt | Financial Soundness | Cash Balance as a fraction of Total Liabilities |
| Cash Flow Margin | cfm | Financial Soundness | Income before Extraordinary Items and Depreciation as a fraction of Sales |
| Short-Term Debt/Total Debt | short.debt | Financial Soundness | Short-term Debt as a fraction of Total Debt |

**Table 2** (continued)

| Financial ratio | Variable name | Category | Formula |
|---|---|---|---|
| Profit Before Depreciation/Current Liabilities | profitjet | Financial Soundness | Operating Income before D &A as a fraction of Current Liabilities |
| Current Liabilities/Total Liabilities | curr_debt | Financial Soundness | Current Liabilities as a fraction of Total Liabilities |
| Total Debt/EBITDA | debt_ebitda | Financial Soundness | Gross Debt as a fraction of EBITDA |
| Long-term Debt/Book Equity | dltt_be | Financial Soundness | Long-term Debt to Book Equity |
| Interest/Average Long-term Debt | int_debt | Financial Soundness | Interest as a fraction of average Long-term debt based on most recent two periods |
| Interest/Average Total Debt | intjotdebt | Financial Soundness | Interest as a fraction of average Total Debt based on most recent two periods |
| Long-term Debt/Total Liabilities | lt_debt | Financial Soundness | Long-term Debt as a fraction of Total Liabilities |
| Total Liabilities/Total Tangible Assets | lt_ppent | Financial Soundness | Total Liabilities to Total Tangible Assets |
| Cash Conversion Cycle (Days) | cash_conversion | Liquidity | Inventories per daily COGS plus Account Receivables per daily Sales minus Account Payables per daily COGS |
| Cash Ratio | cash_ratio | Liquidity | Cash and Short-term Investments as a fraction of Current Liabilities |
| Current Ratio | curr_ratio | Liquidity | Current Assets as a fraction of Current Liabilities |
| Quick Ratio (Acid Test) | quick_ratio | Liquidity | Quick Ratio: Current Assets net of Inventories as a fraction of Current Liabilities |
| Accruals/Average Assets | Accrual | Other | Accruals as a fraction of average Total Assets based on most recent two periods |
| Research and Development/Sales | RD.SALE | Other | R &D expenses as a fraction of Sales |
| Avertising Expenses/Sales | adv_sale | Other | Advertising Expenses as a fraction of Sales |
| Labor Expenses/Sales | staff_sale | Other | Labor Expenses as a fraction of Sales |
| Effective Tax Rate | efftax | Profitability | Income Tax as a fraction of Pretax Income |
| Gross Profit/Total Assets | GProf | Profitability | Gross Profitability as a fraction of Total Assets |
| After-tax Return on Average Common Equity | aftret_eq | Profitability | Net Income as a fraction of average of Common Equity based on most recent two periods |
| After-tax Return on Total Stockholders' Equity | aftret_equity | Profitability | Net Income as a fraction of average of Total Shareholders' Equity based on most recent two periods |

**Table 2** (continued)

| Financial ratio | Variable name | Category | Formula |
|---|---|---|---|
| After-tax Return on Invested Capital | aftretjnvcapx | Profitability | Net Income plus Interest Expenses as a fraction of Invested Capital |
| Gross Profit Margin | gpm | Profitability | Gross Profit as a fraction of Sales |
| Net Profit Margin | npm | Profitability | Net Income as a fraction of Sales |
| Operating Profit Margin After Depreciation | opmad | Profitability | Operating Income After Depreciation as a fraction of Sales |
| Operating Profit Margin Before Depreciation | opmbd | Profitability | Operating Income Before Depreciation as a fraction of Sales |
| Pre-tax Return on Total Earning Assets | pretret_earnat | Profitability | Operating Income After Depreciation as a fraction of average Total Earnings Assets (TEA) based on most recent two periods, where TEA is defined as the sum of Property Plant and Equipment and Current Assets |
| Pre-tax return on Net Operating Assets | pretret_noa | Profitability | Operating Income After Depreciation as a fraction of average Net Operating Assets (NOA) based on most recent two periods, where NOA is defined as the sum of Property Plant and Equipment and Current Assets minus Current Liabilities |
| Pre-tax Profit Margin | ptpm | Profitability | Pretax Income as a fraction of Sales |
| Return on Assets | roa | Profitability | Operating Income Before Depreciation as a fraction of average Total Assets based on most recent two periods |
| Return on Capital Employed | roce | Profitability | Earnings Before Interest and Taxes as a fraction of average Capital Employed based on most recent two periods, where Capital Employed is the sum of Debt in Long-term and Current Liabilities and Common/Ordinary Equity |
| Return on Equity | roe | Profitability | Net Income as a fraction of average Book Equity based on most recent two periods, where Book Equity is defined as the sum of Total Parent Stockholders' Equity and Deferred Taxes and Investment Tax Credit |
| Total Debt/Equity | de_ratio | Solvency | Total Liabilities to Shareholders' Equity (common and preferred) |
| Total Debt/Total Assets | debt_assets | Solvency | Total Debt as a fraction of Total Assets |
| Total Debt/Total Assets | debt_at | Solvency | Total Liabilities as a fraction of Total Assets |

**Table 2** (continued)

| Financial ratio | Variable name | Category | Formula |
|---|---|---|---|
| Total Debt/Capital | debt_capital | Solvency | Total Debt as a fraction of Total Capital, where Total Debt is defined as the sum of Accounts Payable and Total Debt in Current and Long-term Liabilities, and Total Capital is defined as the sum of Total Debt and Total Equity (common and preferred) |
| After-tax Interest Coverage | intcov | Solvency | Multiple of After-tax Income to Interest and Related Expenses |
| Interest Coverage Ratio | intcov_ratio | Solvency | Multiple of Earnings Before Interest and Taxes to Interest and Related Expenses |
| Dividend Payout Ratio | dpr | Valuation | Dividends as a fraction of Income Before Extra. Items |
| Forward P/E to 1-year Growth (PEG) ratio | PEG_1yrforward | Valuation | Price-to-Earnings, excl. Extraordinary Items (diluted) to 1-Year EPS Growth rate |
| Forward P/E to Long-term Growth (PEG) ratio | PEG1tgforward | Valuation | Price-to-Earnings, excl. Extraordinary Items (diluted) to Long-term EPS Growth rate |
| Trailing P/E to Growth (PEG) ratio | PEGjxailing | Valuation | Price-to-Earnings, excl. Extraordinary Items (diluted) to 3-Year past EPS Growth |
| Book/Market | bm | Valuation | Book Value of Equity as a fraction of Market Value of Equity |
| Shillers Cyclically Adjusted P/E Ratio | capei | Valuation | Multiple of Market Value of Equity to 5-year moving average of Net Income |
| Dividend Yield | divyield | Valuation | Indicated Dividend Rate as a fraction of Price |
| Enterprise Value Multiple | evm | Valuation | Multiple of Enterprise Value to EBITDA |
| Price/Cash flow | pcf | Valuation | Multiple of Market Value of Equity to Net Cash Flow from Operating Activities |
| P/E (Diluted, Excl. EI) | pe.exi | Valuation | Price-to-Earnings, excl. Extraordinary Items (diluted) |
| P/E (Diluted, Incl. EI) | pejnc | Valuation | Price-to-Earnings, incl. Extraordinary Items (diluted) |
| Price/Operating Earnings (Basic, Excl. EI) | pe_op_basic | Valuation | Price to Operating EPS, excl. Extraordinary Items (Basic) |
| Price/Operating Earnings (Diluted, Excl. EI) | pe_op_dil | Valuation | Price to Operating EPS, excl. Extraordinary Items (Diluted) |
| Price/Sales | ps | Valuation | Multiple of Market Value of Equity to Sales |
| Price/Book | ptb | Valuation | Multiple of Market Value of Equity to Book Value of Equity |

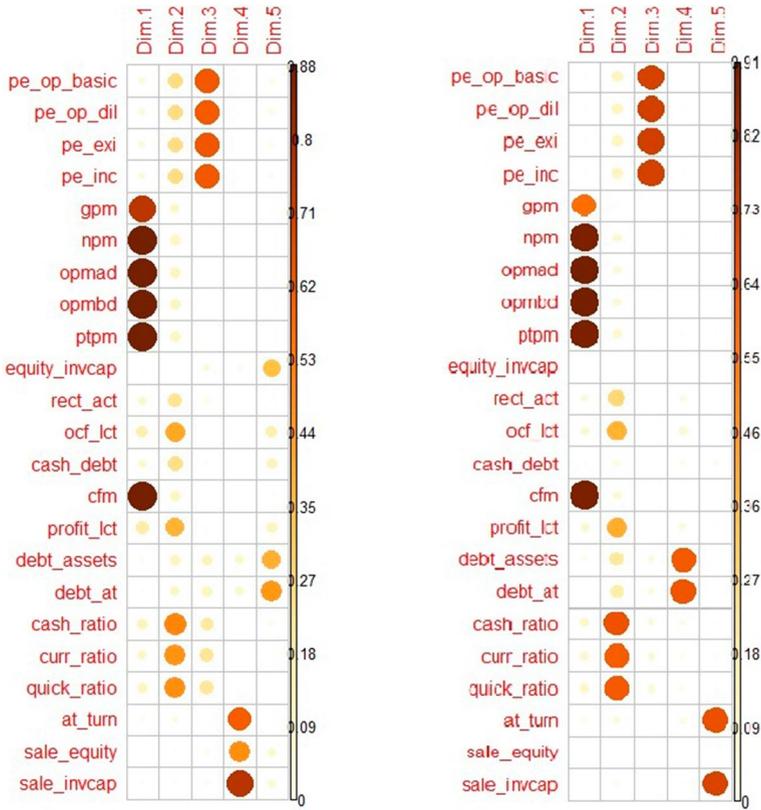

**Fig. 11** Screening plot: the figure shows the results after the first screening (Left panel for "before" COVID-19 and right panel for "after" COVID-19). The size and color of the circles represent the strength of the correlation between the financial ratios (rows) and principal components (columns). Table 2 contains the full explanation of abbreviations. *Data source* Financial ratios, WRDS database. (Color figure online)

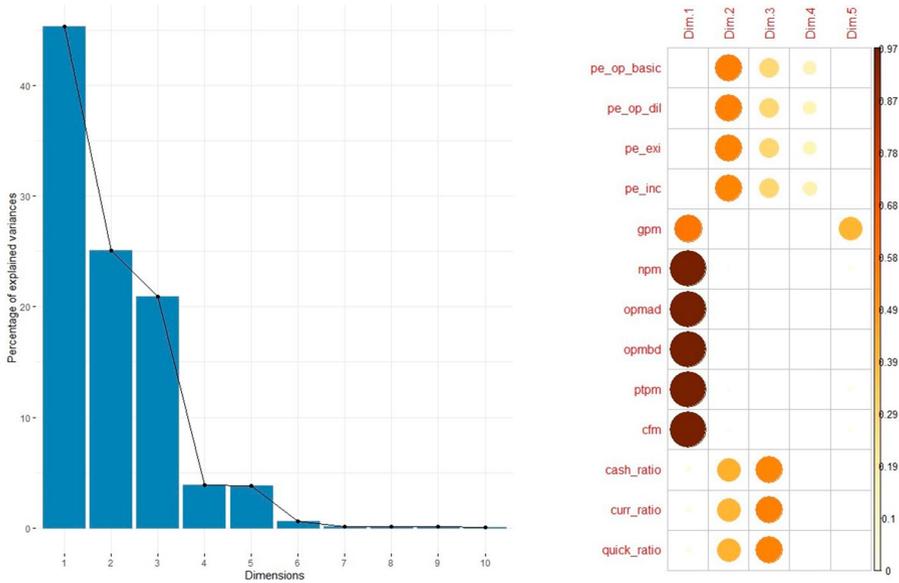

**Fig. 12** Scree and correlation plot: the figure shows the results of PCA at the second step after the COVID-19 outbreak. The left panel (scree plot) shows the proportion of variation explained by each component. The right panel (correlation plot) presents the correlation of each financial ratio with the corresponding dimension (component). The size and color of the circles represent the strength of the correlation between the financial ratios (rows) and principal components (columns). Table 2 contains the full explanation of the abbreviation. *Data source* Financial ratios, WRDS database. (Color figure online)

**Funding** Open access funding provided by Università degli Studi di Napoli Federico II within the CRUI-CARE Agreement.

# References

Accorigi, A. 2011. Energy and climate policy, local production of renewable energy and resilience: A case study on the Ouse valley energy service company, University of Essex (mimeo).

Acharya, V.V., and S. Steffen. 2020. The risk of being a fallen angel and the corporate dash for cash in the midst of COVID. *Review of Corporate Finance Studies* 9 (3): 30–471. https://doi.org/10.1093/rcfs/cfaa013.

Acharya, V.V., S. Lenzu, and O. Wang. 2021b. Zombie lending and policy traps. NBER Working Paper No. 29606.

Albuquerque, R.A., Y. Koskinen, S. Yang, and C. Zhang. 2020. Love in the time of COVID-19: The resiliency of environmental and social stocks. SSRN 3583611.

Alfaro, L., A. Chari, A. Greenland, and P. Schott. 2020. Aggregate and firm-level stock returns during pandemics, in real time. NBER Working Papers 26950. Cambridge: National Bureau of Economic Research.

Barry, J.W., M. Campello, J.R. Graham, Y. Ma, and Y. 2022. Corporate flexibility in a time of crisis. *Journal of Financial Economics* 144: 780–806.

Blickle, K., and J.A.C. Santos. 2022. The costs of corporate debt overhang. SSRN 3708502.

Bordalo, P., N. Gennaioli, R. La Porta, and A. Shleifer. 2020. Expectations of fundamentals and stock market puzzles. Technical report, Working paper.

Botosan, C.A., and M.A. Plumlee. 2005. Assessing alternative proxies for the expected risk premium. *The Accounting Review* 80 (1): 21–53.

Bunting, S., N. Kundu, and N. Ahmed. 2011. Rice-shrimp farming ecocultures in the Sundarbans of Bangladesh and West Bengal, India. University of Essex (mimeo).

Charlot, S., and G. Duranton. 2004. Communication externalities in cities. *Journal of Urban Economics* 56 (3): 581–613.

Cheema-Fox, A., B.R. LaPerla, G. Serafeim, and H. Wang. 2020. Corporate resilience and response during COVID-19. *Journal of Applied Corporate Finance* 33 (2): 24–40.

Cejnek, G., O. Randl, and J. Zechner. 2021. The COVID-19 pandemic and corporate dividend policy. *Journal of Financial and Quantitative Analysis* 56: 2389–2410.

Chen, L., and X. Zhao. 2009. Return decomposition. *Review of Financial Studies* 22: 5213–5249.

Crouzet, N., and F. Tourre. 2021. Can the cure kill the patient? Corporate credit interventions and debt overhang. SSRN 3954581.

Daadmehr, E. 2022. Resilience and asset pricing in COVID-19 disaster. SSRN 4288219.

Daadmehr, E. 2023. COVID intensity, resilience and expected returns. SSRN 4664619.

De la O, R., and S. Myers. 2020. Subjective cash flow and discount rate expectations. *Journal of Finance* 76 (3): 1339–1387.

Ding, W., R. Levine, C. Lin, and W. Xie. 2021. Corporate immunity to the COVID-19 pandemic. *Journal of Financial Economics* 141 (2): 802–830.

Dingel, J.I., and B. Neiman. 2020. How many jobs can be done at home? *Journal of Public Economics* 189: 104–235.

Engelberg, J., R.D. Mclean, and J. Pontiff. 2018. Anomalies and news. *The Journal of Finance* 73: 1971–2001.

Fahlenbrach, R., K. Rageth, and R.M. Stulz. 2021. How valuable is financial flexibility when revenue stops? Evidence from the Covid-19 crisis. *Review of Financial Studies* 34: 5474–5521.

Glossner, S., P. Matos, S. Ramelli, and A.F. Wagner. 2020. Do institutional investors stabilize equity markets in crisis periods? Evidence from Covid-19. Swiss Finance Institute Research Paper No. 20-56. European Corporate Governance Institute - Finance Working Paper No. 688/2020. SSRN 3655271.

Happ, C., and S. Greven. 2018. Multivariate functional principal component analysis for data observed on different (dimensional) domains. *Journal of the American Statistical Association* 113 (522): 649–659. https://doi.org/10.1080/01621459.2016.1273115.

Happ-Kurz, C. 2020. Object-oriented software for functional data. *Journal of Statistical Software* 93 (5): 1–38.

Hensvik, L., T. Le Barbanchon, and R. Rathelot. 2020. Which jobs are done from home? Evidence from the American Time Use Survey. CEPR Discussion Paper No. DP14611. SSRN 3594233.

Hua, J., L. Peng, R.A. Schwartz, and N.S. Alan. 2020. Resiliency and stock returns. *The Review of Financial Studies* 33 (2): 747–782.

Kalemli-Ozcan, S., L. Laeven, and D. Moreno. 2022. Debt overhang, rollover risk, and corporate investment: Evidence from the European crisis. *Journal of the European Economic Association* 20 (6): 2353–2395.

Koren, M., and R. Peto. 2020. Business disruptions from social distancing. *PLoS ONE* 15 (9): e0239113. https://doi.org/10.1371/journal.pone.0239113.

Kothari, S.P., E. So, and R. Verdi. 2016. Analysts' forecasts and asset pricing: A survey. *Annual Review of Financial Economics* 8 (1): 197–219.

Landier, A., and D. Thesmar. 2020. Earnings expectations in the COVID Crisis. HEC Paris Research Paper No. FIN-2020-1377.

Li, K., X. Liu, F. Mai, and T. Zhang. 2021. The role of corporate culture in bad times: Evidence from the COVID-19 pandemic. *Journal of Financial and Quantitative Analysis* 56 (7): 2545–2583.

Lochster, L.A., and P.C. Tetlock. 2020. What drives anomaly returns? *The Journal of Finance* 75: 1417–1455.

Ohlson, J.A., and B.E. Juettner-Nauroth. 2005. Expected EPS and EPS growth as determinants of value. *Review of Accounting Studies* 10: 349–365.

Pagano, M., C. Wagner, and J. Zechner. 2023. Disaster resilience and asset prices. *Journal of Financial Economics*. https://doi.org/10.2139/ssrn.3603666.

Pagano, M., and J. Zechner. 2022. COVID-19 and corporate finance. *The Review of Corporate Finance Studies* 11 (2): 849–879.

Papanikolaou, D., and L.D.W. Schmidt. 2022. Working remotely and the supply-side impact of COVID-19. *The Review of Asset Pricing Studies* 12: 53–111.

Pettenuzzo, D., R. Sabbatucci, and A. Timmermann. 2022. Firm value and payout suspensions during financial market distress. Swedish House of Finance Research Paper No. 21-10. SSRN 3823258.

Ramelli, S., and A.F. Wagner. 2020. Feverish stock price reactions to covid-19. *Review of Corporate Finance Studies* 9 (3): 622–655.

Sedlmayer, A., and S. Boehm. 2011. Alternative agri-food networks in the Colchester area and their contribution to developing local resilience. University of Essex (mimeo).

Tian, L. 2019. Division of labor and productivity advantage of cities: Theory and evidence from Brazil. CEPR Discussion Paper No. DP16590. SSRN 3960170.

Walker, B., C. Holling, S. Carpenter, and A. Kinzig. 2004. Resilience, adaptability and transformability in social–ecological systems. *Ecology and Society* 9 (2): 5.

Wang, J., J. Yang, B. Iverson, and R. Kluender. 2021. Bankruptcy and the COVID-19 crisis. Harvard Business School Working Paper No. 21-041.